\documentclass[10pt]{iopart}

\usepackage{iopams} 

\expandafter\let\csname equation*\endcsname\relax
\expandafter\let\csname endequation*\endcsname\relax 

\usepackage{amssymb}
\usepackage{amsthm}
\usepackage{MnSymbol}
\usepackage{graphicx}
\usepackage{dcolumn}
\usepackage{bm}
\usepackage{chemarrow}
\usepackage{graphicx}
\usepackage{epsfig}
\usepackage{mathrsfs}
\usepackage{color}

\newcommand{\ua}{\uparrow}
\newcommand{\da}{\downarrow}
\newcommand{\be}{\begin{eqnarray}}
\newcommand{\ee}{\end{eqnarray}}
\newcommand{\bea}{\begin{eqnarray}}
\newcommand{\eea}{\end{eqnarray}}
\newcommand{\bes}{\begin{subequations}\bea}
\newcommand{\ees}{\eea\end{subequations}}
\newcommand{\ba}{\begin{array}}
\newcommand{\ea}{\end{array}}

\newcommand{\boj}{\boldsymbol{j}}
\newcommand{\boa}{\boldsymbol{a}}
\newcommand{\bol}{\boldsymbol{\ell}}
\newcommand{\boc}{\boldsymbol{c}}
\newcommand{\boe}{\boldsymbol{e}}
\newcommand{\boL}{\mathbf{L}}
\newcommand{\boE}{\mathbf{E}}
\newcommand{\boB}{\mathbf{B}}
\newcommand{\boeta}{\boldsymbol{\eta}}

\newcommand{\tc}[1]{\textcolor{black}{#1}}

\newtheorem{theorem}{Theorem}
\newtheorem{definition}{Definition}

\begin{document}
 
\title[System/environment duality of nonequilibrium network observables]{System/environment duality \\ of nonequilibrium network observables}

\author{Matteo Polettini$^1$}

\address{$^1$ Facult\'e des Sciences, de la Technologie et de la Communication
162 A, avenue de la Fa\"\i{}encerie, L-1511 Luxembourg (Grand Duchy of Luxembourg) }

\ead{matteo.polettini@uni.lu}

\begin{abstract} On networks representing probability currents between states of a system, we generalize Schnakenberg's theory of nonequilibrium observables to nonsteady states, with the introduction of a new set of macroscopic observables that, for planar graphs, are related by a duality. We apply this duality to the linear regime, obtaining a dual proposition for the minimum entropy production principle, and to discrete electromagnetism, finding that it exchanges fields with sources. We interpret duality as reversing the role of system and environment, and discuss generalization to nonplanar graphs. \tc{The results are based on two theorems regarding the representation of bilinear and quadratic forms over the edge vector space of an oriented graph in terms of observables associated to cycles and cocycles.}
\end{abstract}



\section{\label{intro}Introduction}

In a seminal paper \cite{schnak}, J. Schnakenberg engaged in the definition of the  fundamental macroscopic observables of NonEquilibrium Statistical Mechanics, grossly conceived as a theory of the internal flows of a system. \tc{The construction is rooted in algebraic graph theory, where the graph represents the discrete state space of the system and edges represent possible transitions between states.} In accordance with our modern understanding of Quantum Mechanics and Quantum Field Theory, where adiabatic phases  and Wilson loops play an ever more prominent role, he interpreted circuitations  of certain variables as the  constraints which prevent a system from relaxing to equilibrium.  Born out of the study of biophysical systems \cite{hill,schnak2}, and recently finding growing applications to chemical reaction networks, molecular motors and transport phenomena \cite{polettiniespo,andrieuxA,andrieuxB,liepeltA,liepeltB,faggionato,andrieux2A,andrieux2B,qiansA,qiansB}, his analysis has a deep geometrical and combinatorial content \cite{qiansbook,kalpa,polettinigauge}. It is the backbone for the comprehension of Non-Equilibrium Steady States (NESSs) \cite{qiansbook,zia}, to which the theory is so far restricted. 

The aim of this contribution is to go beyond NESSs, generalizing Schnakenberg's construction to arbitrary states. The complete theory of nonequilibrium observables turns out to enjoy a duality which exchanges forces with currents, the concept of steadiness with that of detailed-balancing of the external constraints. While steadiness is a property of the state of the system, detailed-balancing of the external forces is a property of the state of the environment: whence this contribution's title. \tc{Mathematically, the result is based on a decomposition of bilinear forms (e.g. the so-called {\it entropy production}) and of quadratic forms (e.g. entropy production in the linear regime) defined over the edge vector space of an oriented graph in terms of quantities associated to a basis of cycles and of cocycles constructed starting from a spanning tree of the graph.}

In Sec.\,\ref{schnak1} we give a simple example of our construction. In Sec.\,\ref{schnak} we review Schnakenberg's theory, recast it in graph-theoretical terms and generalize his theorem on the steady entropy production to non-steady states. In Sec.\,\ref{duality} we discuss duality. In  Sec.\,\ref{linear} we explore the linear regime, proving a second theorem on the representation of the entropy production which allows to derive the minimum entropy production principle and its dual proposition. In Sec.\,\ref{em}, to show the generality of our theorem, as an exercise we apply it to electromagnetism on a lattice, comparing our duality with the electromagnetic duality. We draw conclusions in Sec.\,\ref{concl}.

\tc{\section{\label{schnak1}Simple example}}

Schnakenberg's focus was on \tc{Markovian master equations of the kind}
\bea \dot{\rho}_v(t) = \sum_{v'} \big[ w_{vv'} \rho_{v'}(t) -  w_{v'v} \rho_{v}(t)\big], \label{eq:rate}  \eea
\tc{where vertex $v$ belongs to a finite state space $V$ of the system, $\rho:[0,\infty)\times V \to [0,1]$ is a normalized probability density, differentiable with respect to time $t$, and $w:V\times V \to [0,\infty)$ are positive transition rates along edges $e =v \gets v'$ of a network, or oriented graph, $G$. Let us introduce the {\it mesoscopic} currents and forces as the edge variables (dropping the explicit time dependencies)}
{\color{black}
\bes
j_{vv'} & = &  w_{vv'} \rho_{v'}  -  w_{v'v} \rho_{v} \\
a_{vv'} & = &  \ln \frac{w_{vv'} \rho_{v'} }{ w_{v'v} \rho_v}.
\ees
Notice that they are antisymmetric by inversion of the orientation of the edge, $j_{vv'}=- j_{v'v}$, $a_{vv'}=-a_{v'v}$. The major insight of Schnakenberg was to identify the total entropy produced by a system governed by such a master equation with the following bilinear form} 
\bea \sigma = \sum_{v,v'} w_{vv'} \rho_{v'} \ln \frac{w_{vv'} \rho_{v'} }{ w_{v'v} \rho_v} = \frac{1}{2}\sum_{v,v'} j_{vv'} a_{vv'}. \label{eq:EP1} \ee
\tc{Furthermore}, he realized that at a NESS, that is, when Kirchhoff's law $\dot{\rho}= 0$ is satisfied, currents can be expressed as linear combinations of a certain number of \emph{macroscopic internal} currents $J_\alpha$, which flow along some preferred edges of the graph, later to be identified. On the $\dot{\rho}=0$ shell, the entropy production comes down to $\sigma = J_\alpha A^\alpha$ (repeated indices are implicitly summed over), where the conjugate variables $A^\alpha$ are seen to be circulations of the mesoscopic forces around suitable cycles
\bea A^\alpha = \ln \frac{ w_{v_1 v_2} w_{v_2 v_3} \ldots w_{v_n v_1}}{w_{v_1 v_n} \ldots w_{v_3 v_2}w_{v_2 v_1}},  \ee
\tc{and the index $\alpha$ ranges over a complete set of cycles (see below)}. A system whose steady state $\rho^{ss}$ makes all mesoscopic currents and forces vanish, $w_{vv'}\rho^{ss}_{v'}= w_{v'v}\rho^{ss}_{v}$, is said to satisfy detailed balance. Schnakenberg's choice of circuitations as the  fundamental observables, indicative of the nonequilibrium nature of the system, is motivated by the well-known fact that they all vanish if and only if  the steady state is detailed balanced (Kolmogorov's criterion) \cite{zia}. Moreover, they do not depend on the system's macrostate $\rho$: they are external constraints, which conceptually are imputable to the state of the environment. Hence, in the following, we will refer to  detailed-balanced systems as those which satisfy Kolmogorov's criterion.

A comment on the usage of the scale words is in order. Schnakenberg referred to $j_{vv'}$ as the \emph{microscopic} currents, and to the observables we are going to build as macroscopic. However, later developments in the stochastic thermodynamics of master equation systems (see \cite{seifert} and references therein) allow to identify single-trajectory analogs of thermodynamical quantities, such as currents and entropy production, whose averages over paths return $j_{vv'}, \sigma,$ etc. This suggests to reserve the word ``microscopic'' for this further layer, and to adopt ``mesoscopic'' for the averaged quantities, irregardless of the spatial dimensions that are involved in the problem.

Indeed, Schnakenberg's analysis can be extended to any graph whose edges bear a couple of antisymmetric conjugate variables, one of which obeys Kirchhoff's Law at the nodes. Thence abandoning master equation thermodynamics ---but retaining the nomenclature, we review and complement Schnakenberg's definitions with a new set of conjugate macroscopic observables.

The results are based on a decomposition theorem of the entropy production in cycles and  flows (or cocycles) of the graph. To give a first hint, consider the $3$-level system depicted with straight lines in fig.\ref{fig:dual}a,
\bea v_1 ~  \autorightleftharpoons{$j_3,a_3$}{$-j_3,-a_3$} ~ v_2 ~ \autorightleftharpoons{$j_1,a_1$}{$-j_1,-a_1$} ~ v_3 ~  \autorightleftharpoons{$j_2,a_2$}{$-j_2,-a_2$} ~  v_1 .   \ee
By the Handshaking lemma ($\sum_v = 2 \sum_e$), the entropy production can be recast as $\sigma =  a_1 j_1 + a_2 j_2 + a_3 j_3$. We reshuffle, add and subtract terms to obtain
\bea \sigma ~=~ \stackrel{A^1}{\overbrace{(a_1 + a_2 + a_3)}}  j_1 +   a_2 \stackrel{J_\ast^2}{\overbrace{(j_2-j_1)}} + \,  a_3\stackrel{J_\ast^3}{\overbrace{(j_3-j_1)}},   \ee
where the overbraces are used to define, along with one Schnakenberg circulation $A^1$  (fig.\ref{fig:dual}b) and its conjugate internal macroscopic current $J_1 = j_1$,  the macroscopic \emph{external} currents $J^2_\ast$, flowing out of vertex $v_3$ (fig.\ref{fig:dual}c), and $J^3_\ast$, flowing into vertex $v_2$ (fig.\ref{fig:dual}d). Since, by  (\ref{eq:rate}), $J^2_\ast =-\dot{\rho}_3$ and $J^3_\ast =\dot{\rho}_2$, it is conceptually appropriate to ascribe these observables to the state of the system.  The vanishing of $A^1$ provides balancing, the vanishing of $J^2_\ast,J^3_\ast$ defines steadiness. In graph-theoretical language, $J^2_\ast$ and $J^3_\ast$ are weighted cocycles, that is, edge sets whose removal disconnects the vertex set $V$ into two noncommunicating components: they measure the total flow from one set towards the other. The asterisk will later be interpreted in terms of duality.

~

\section{\label{schnak}Schnakenberg revisited}

\tc{
Let $G = (V,E,\partial)$ be an oriented connected graph without loops but possibly with multiple edges, with $|V|$ vertices $v \in V$ and $|E|$ edges $e \in E$.
Edges carry an arbitrary orientation (a choice of tip and tail vertices), with $-e$ designating the inverse edge. The topology of the graph is completely described by the incidence matrix $\partial: \mathbb{R}^E \to \mathbb{R}^V$}
\bea \partial_v^e =  \left\{ \begin{array}{ll} + 1, &~~ \mathrm{if}~ \stackrel{e}{\gets} v \\ 
-1,  &~~ \mathrm{if}~ \stackrel{e}{\to} v\\
0, & ~~\mathrm{elsewhere}
\end{array} \right. .  \eea   

We employ an algebraic approach to graph theory \cite{graphs,nakanishi}, working with integer linear combinations of edges in the lattice $\mathcal{E} = \mathbb{Z}^{E}$, upon which $\partial$ acts as a boundary operator. It is a standard result that $\partial$ induces an orthogonal decomposition of $\mathcal{E} =\mathcal{C} \oplus \mathcal{C}^\ast$ into the  cycle space  $\mathcal{C} = ker( \partial)$ and the cocycle space $\mathcal{C}^\ast = rowspace(\partial)$. The dimension of the cycle space is given by the cyclomatic number  $|C| = |E|-|V|+1$, whence by the rank-nullity theorem the cocycle space has dimension $|V|-1$.

\begin{figure}
\hspace{3cm}\includegraphics[scale=0.33]{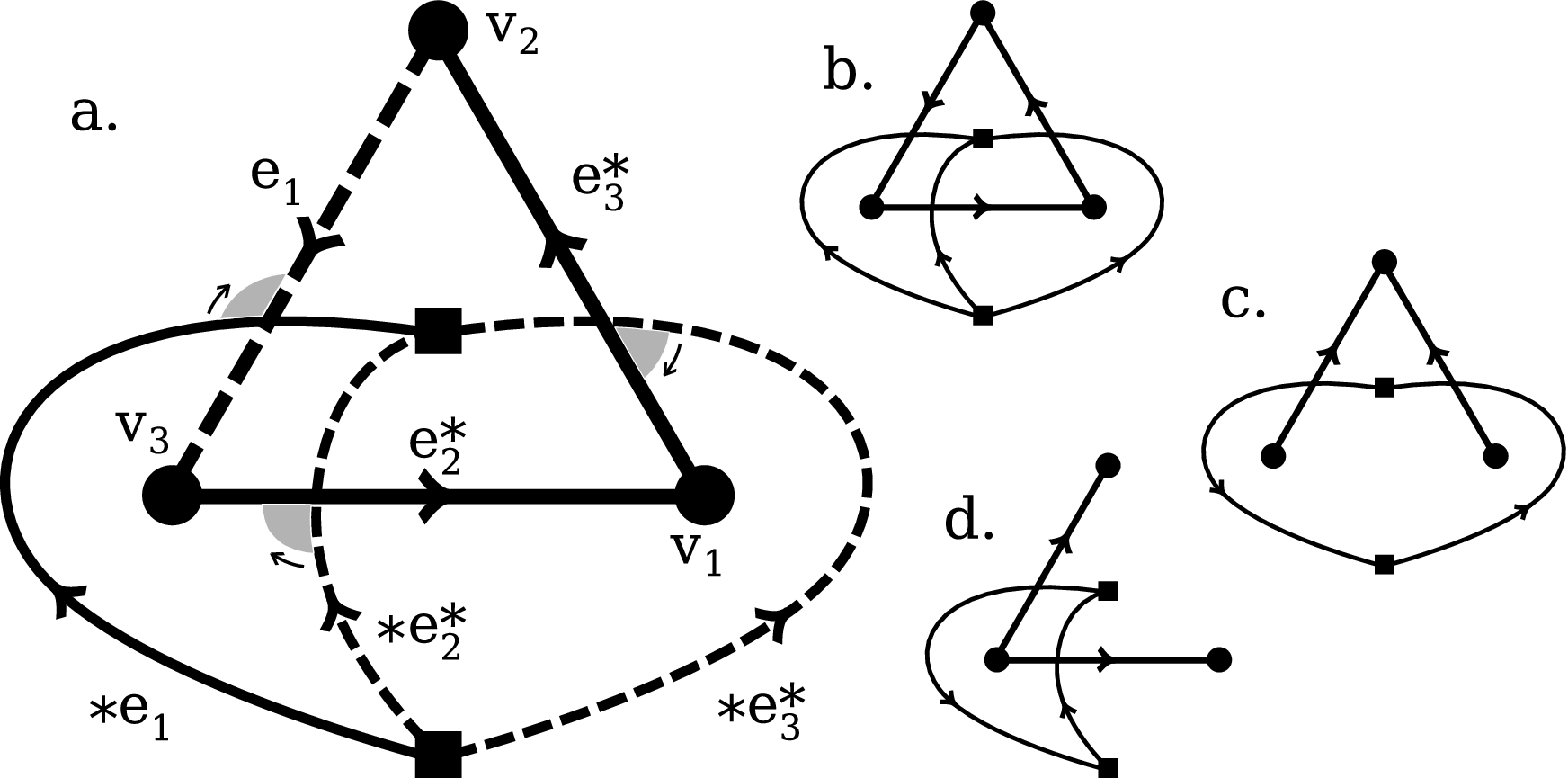}
\caption{\label{fig:dual}(a.) A planar graph and its dual, with vertices depicted as bullets, edges (arbitrarily oriented) by straight lines, dual vertices with boxes, dual edges with curved lines. Spanning trees are solid, their complements are dashed. Shaded angles indicate clockwise orientations. (b.) The oriented cycle generated by $e_1$, and its dual cocycle. (c. and d.) In straight lines, a fundamental set of cocyles, respectively generated by cochords $e^\ast_2$ and $e^\ast_3$; in curved lines, their dual cycles.}
\end{figure}

From a graphical point of view, cycles $\boc$ are chains of oriented edges such that each vertex is the tip and the tail of an equal number of edges (possibly none). A cycle is simple if it is connected, has no crossings or overlapping edges. A simple cycle can exist in two opposite orientations. A simple cocycle $\boc^\ast$ is a collection of edges whose removal disconnects the vertex set into two components; it might carry one of two possible orientations when all edges point from one of the two components, called the source set $\texttt{s}(\boc^\ast)$, towards the other (see fig.\ref{fig:dual}c,d).

Of all possible integral basis of $\mathcal{E}$, we concentrate on  fundamental sets, which are so built. Let $T \subseteq E$ be a spanning tree of the graph (i.e. a maximal subset of $E$ containing no cycles); we call its edges $\boe^\ast_\mu$ the \emph{cochords}. The remaining edges $e_\alpha \in E \setminus T$ are called \emph{chords}. There are $|V|-1$ cochords and $|C|$ chords.
When a chord $\boe_\alpha$ is added a spanning tree, a simple cycle $\boc^\alpha$ is generated, which can be oriented accordingly with $\boe_\alpha$ (see fig.\ref{fig:dual}a,b). The  fundamental set of cycles $C = \{\boc^\alpha\}$ so generated is a basis for $\mathcal{C}$. Similarly, when a cochord  $\boe^\ast_\mu$ is removed, the spanning tree is disconnected into two components, which identify a simple cocycle $\boc_\ast^\mu$, with orientation dictated by $\boe^\ast_\mu$ (see fig.\ref{fig:dual}a,c,d). Again, the  fundamental set of cocycles $\{\boc^\mu_\ast\}$ is a basis for $\mathcal{C}^\ast$.
The crucial peculiarity of  fundamental sets is that no chord is shared by two cycles, and no cochord is shared by two cocycles. Moreover, any of the sets  $\{\boe_\alpha,\boe_\mu^\ast\}$, $\{\boc^\alpha,\boe_\mu^\ast\}$, $\{\boc^\alpha,\boc^\mu_\ast\}$, $\{\boe_\alpha,\boc^\mu_\ast\}$ forms a basis for $\mathcal{E}$.

\tc{
\begin{definition}
We introduce:
\begin{itemize}
\item[(1)] The \emph{mesoscopic currents} $j:E\to\mathbb{R}$, antisymmetric by inversion of the orientation of an edge, $j_{-e} = -j_e$;
\item[(2)] The antisymmetric \emph{mesoscopic forces} $a:E\to\mathbb{R}$;
\item[(3)] A bilinear form called \emph{entropy production} given by
\bea
\sigma = \sum_{e} j_e a_e = (\boj,\boa), \label{eq:EP}
\eea
where in the r.h.s. is the euclidean scalar product on the edge set in shorthand;
\item[(4)] The \emph{macroscopic observables}
\be
J_\alpha = (\boe_\alpha,\boj),  \quad  J_\ast^\mu = (\boc_\ast^\mu,\boj),  \quad
A^\alpha = (\boc^\alpha,\boa), \quad  A_\mu^\ast = (\boe_\mu^\ast,\boa). \label{defs}
\eea 
In left-to-right order: \emph{internal currents} flow along  fundamental chords, \emph{external currents} are the total flow out of the source set of a cocycle, \emph{external forces} are circuitations of forces along the  fundamental cycles, \emph{internal forces} are exerted along edges of the spanning tree. 
\end{itemize}
\end{definition}}

In terms of the incidence matrix,  Kirchhoff's Law reads $\partial \boj = 0$, and the r.h.s. of  (\ref{eq:rate}) reads $\dot{\rho} + \partial \boj=0$. Oriented overlaps between edge sets can be succinctly expressed in terms of the scalar product:
\bea \label{eq:ort}
(\boc^\alpha,\boe_\beta) = \delta^\alpha_\beta, \quad (\boe^\ast_\mu,\boc^\nu_\ast,) = \delta_\mu^\nu, \quad (\boc^\alpha,\boc_\ast^\mu) = 0, \quad (\boe_\mu^\ast,\boe_\alpha) = 0. 
\ee
We are now ready to formulate the core theorem.

\begin{theorem}\label{theorem1}
\tc{Let the entropy production $\sigma = (\boj,\boa)$ be a bilinear form over the edge space of a connected oriented graph without loops. Consider an arbitrary spanning tree of the graph and let $\boc^\alpha$ be the basis of oriented cycles generated by the chords $\boe_\alpha$ of the spanning tree, and $\boc^\nu_\ast$ be the basis of oriented cocycles generated by the cochords $\boe^\ast_\mu$. Then the entropy production can be decomposed into a steady-state and a detailed-balanced term, $\sigma = \sigma_{ss} + \sigma_{db}$, given respectively by
\bea
\sigma_{ss} = A^\alpha J_\alpha, \qquad\sigma_{db} = J_\ast^\mu A_\mu^\ast  
\eea
\tc{where the macroscopic observables are defined in Eq.(\ref{defs})}.}
\end{theorem}

\begin{proof}
The strategy is to find the general solution to the continuity equation with sources $\dot{\rho} +\partial \boj = 0$. 
Here $\dot{\rho} \in \mathbb{R}^{V}$ is the current injected at the vertices, and it is constrained to satisfy $\sum_{v \in V} \dot{\rho}_v = 0$. Since any $|V|-1$ rows of $\partial$ span the cocycle space, $\dot{\rho}_v$ is expressible as a linear combination of a  fundamental set of external currents, and vice versa. One can easily show that
\bea -\dot{\rho}^{\, \mu} = J_\ast^\mu   = -\sum_{v \in \texttt{s}(\boc_\ast^\mu)} \dot{\rho}_v. \label{eq:sourceset}  \eea
\tc{The reasoning is the following. Consider two cocycles $c^\ast_1$ and $c^\ast_2$ emanating from two disjoint source sets $S_1$ and $S_2$. We want to know the composition of $c^\ast_1+c^\ast_2$. If two vertices $v_1 \in S_1$ and $v_2\in S_2$ are neighbors, then the edge $v_1\gets v_2$ will occur both in $c^\ast_1$ and $c^\ast_2$ with opposite orientation, hence canceling out in $c^\ast_1+c^\ast_2$. What remains is the set of edges that emanate from $S_1\cup S_2$ towards vertices in $V\setminus(S_1\cup S_2)$, which is precisely the cocycle emanating from $S_1\cup S_2$. Hence, since individual vertices are disjoint, Eq.\,(\ref{eq:sourceset}) is proven.} Physically: the flow out of a source set is equal to (minus) the sum of the injected currents within the set.
The general solution can be found as a particular solution plus the general solution of the homogeneous equation associated to it. Solving $\partial \boj=0$ yields a superposition of cycles $\sum_\alpha \lambda_\alpha \boc^\alpha$. As to the particular solution, since $\{\boc^\alpha,\boe_\mu^\ast\}$ is a basis for $\mathcal{E}$, we can tune the cycle currents so as to make  currents along chords vanish. We then only need to specify the particular solution along cochords, obtaining
\bea
\boj = \lambda_\alpha \boc^\alpha + \lambda_\ast^\mu \boe_\mu^\ast. \label{eq:gensol} 
\eea
Inserting (\ref{eq:gensol}) into the definitions (\ref{defs}), and using the orthonormality relations (\ref{eq:ort}), we identify $J_\alpha =  \lambda_\alpha$, and $J_\ast^\mu = \lambda_\ast^\mu$. Further insertion into (\ref{eq:EP}) yields our thesis.\end{proof}

\tc{
\begin{definition}A \emph{detailed-balanced} system has vanishing external macroscopic forces $A^\alpha = 0$, in which case $\sigma_{ss}$  vanishes for all values of the internal currents. A \emph{steady state} has vanishing external macroscopic currents $J_\ast^\mu = 0$, in which case $\sigma_{db}$ vanishes for all values of the internal forces. When both vanish we talk of \emph{equilibrium states}.
\end{definition}}

\section{\label{duality}Duality}

\tc{If the graph is planar it admits a dual. Then, cycles and cocycles, chords and cochords are dual one to each other. In this section we look at the consequences of duality for our theory and discuss the limitations posed by planarity. We leave further discussion of the physical interpretation to the Conclusions.}

A graph is planar if it can be drawn on the surface of a sphere with non-intersecting edges. Planar embeddings have faces  $f \in F$, i.e.  open neighbors of the sphere which cannot be path-connected without crossing an edge. Their number $|F|=C+1$, including the ``outer'' face, is prescribed by Euler's formula.

The dual graph $G^\ast=(V^\ast,E^\ast,\partial^\ast)$ has one vertex per face, $V^\ast = F$, two dual vertices being connected by one dual edge $\ast e$ per each boundary edge $e$ that the corresponding faces share, so that $E^\ast = E$. Pictorially, after puncturing and flattening the sphere, one will draw a vertex inside each face and a dual edge $\ast e$ crossing $e$, then assign an orientation by clockwise rotating $\ast e$ until it overlaps, tip and tail, with $e$ (see curved lines and shadings in fig.\ref{fig:dual}a). Crucial facts about duality are:
\begin{itemize}
\item[(i)] Up to a reorientation $E\to -E$, it is involutive; 
\item[(ii)] Different embeddings might yield non-isomorphic duals (with different incidence relations);
\item[(iii)] It maps a spanning tree $T$ to the complement $T^\ast= E \setminus T_\ast$ of a spanning tree $T_\ast \subseteq E^\ast$, in such a way that the  fundamental sets generated by $T_\ast$ are the duals of the  fundamental sets generated by $T$, according to the scheme (see fig.\ref{fig:dual}b,c,d)
\bea
\mathrm{chords}  &\leftrightarrow& \mathrm{cochords},  \nonumber \\
\mathrm{cycles}  &\leftrightarrow& \mathrm{cocycles}. \nonumber
\eea
\end{itemize}
Duality can then be applied to the graphical structure of nonequilibrium observables.
So, for example, the map $\boa \leftrightarrow \boj$ leaves $\sigma$ invariant, but switches macroscopic observables with those of the dual graph, mapping internal forces to internal currents and external currents to external forces:
\bea
A^\alpha \leftrightarrow J^\mu_\ast, \quad A_\mu^\ast   \leftrightarrow  J_\alpha, \quad \sigma_{ss} \leftrightarrow \sigma_{db}.  
\eea
Since we ascribed $A^\alpha$ to the state of the environment and $J^\mu_\ast$ to that of the system, it is fair to dub this \emph{system-environment} duality.  Steady states, for which the macroscopic external currents vanish, are dual to detailed-balanced systems, for which the macroscopic external forces vanish: the former are in fact properties of the system under given environmental conditions, while the second are properties of the environment's influence on the system, independently of the system's state. 

Out of the $\boa \leftrightarrow \boj$ special case, we stress that duality is a graph-theoretical property: it tells how well-behaved observables look like from the point of view of the environment and of the system, not which mesoscopic variables enter the construction.

Planarity seems to be a major limitation to the generality of system/environment duality. We argue that this is not the case. Property (iii) listed above is independent of the particular embedding chosen. Indeed, generalizing the concept of a graph to that of an abstract matroid \cite{matroids,sokal}, it turns out that matroids always have a well-defined dual which satisfies property (iii), even though dual matroids might not be visualizable as graphs. In other words, trees and cotrees, cycles and cocycles, chords and cochords always have mutual properties, even when there exists no dual graph.

\section{\label{linear}Linear regime and minimum entropy production}

One major clue that led Schnakenberg to identify chords and cycles as good thermodynamic observables is the fact that, in the linear regime, Onsager's reciprocity relations arise. By ``linear regime'' it is meant that mesoscopic currents and forces satisfy Ohm's law
\bea\boa = \bol \boj + \Or(\boj^2),\eea
where $\bol = \mathrm{diag}\{\ell_1,\ldots,\ell_{|E|}\}$ is a local linear response matrix, connecting mesoscopic quantities edge-by-edge. Suppose that a system, initially at equilibrium, is perturbed to a nearby nonequilibrium steady state. Schnakenberg furnished the macroscopic linear relation $A^\alpha = L^{\alpha\beta} J_\beta$, with a symmetrical linear response matrix $\boL$. In our algebraic formalism, the derivation is straightforward:
\bea
A^\alpha = (\boc^\alpha,\bol \boj) = (\boc^\alpha,\bol \boc^\beta) J_\beta = L^{\alpha\beta} J_\beta .
\eea
The linear response matrix is a weighted superposition of cycles. For master equation systems, this insight is complemented by Andrieux and Gaspard's proof of a Green-Kubo-type of formula for $\boL$ \cite{andrieuxA}.  Let us now linearly perturb an equilibrium state into a nonsteady, but still detailed-balanced configuration. While Kirchhoff's law implies steadiness, detailed balancing follows from the dual relation to Kirchhoff's law, namely $\partial^\ast \boa = 0$. Its solution is by $\boa = A_\mu^\ast \boc_\ast^\mu$. Then
\bea J^\mu_\ast = (\boc_\ast^\mu,\bol^{-1}\boa) =  (\boc_\ast^\mu,\bol^{-1}\boc_\ast^\nu) A_\nu^\ast =  L_\ast^{\mu\nu}A_\nu^\ast   \label{eq:detballin}  \eea  
and the dual response matrix $\boL_\ast$ is a weighted superposition of cocycles. Both matrices $\boL$ and $\boL_\ast$ are symmetric, and under $\bol \leftrightarrow \bol^{-1}$ they are dual one to the other. Similar matrices are employed in electrical circuit analysis \cite{iyer} and in the parametric formulas for Feynman diagrams (see \cite[\S18.4]{bjorken} and \cite[\S3]{nakanishi}). In this contest planar-graph duality has been related to duality between momentum and position representations \cite{david}. Possibly, the most interesting property of $\boL$ and $\boL_\ast$ is that their determinants, which are always nonnull but for very trivial graphs, are independent of the fundamental sets chosen, obey the relation $\det \boL/\det \boL^\ast = \det \bol$, and are related to the $0$-state Potts-model partition function \cite{potts,sokal}.

Another crucial fact is that when the equilibrium state is linearly perturbed in an unconstrained manner (neither into a steady state nor into a detailed balanced configuration), the entropy production can still be written as a block-diagonal bilinear form of the external observables according to the following theorem.

\begin{theorem}\label{theorem2}
{\color{black}Under the assumptions of Theorem \ref{theorem1}, letting $\boa = \bol \boj$, $\bol$ a diagonal invertible matrix, the quadratic entropy production can be decomposed as
\bea
\sigma = (\mathbf{L}^{-1})_{\alpha\beta} A^{\alpha}A^{\beta} + (\mathbf{L}^{-1}_\ast)_{\mu\nu} J_\ast^\mu J_\ast^\nu\label{eq:complete} 
\eea
where the linear response matrices are given by
\bes
L_\ast^{\mu\nu} = (\boc_\ast^\mu,\bol^{-1}\boc_\ast^\nu) A_\nu^\ast  \\
L^{\alpha\beta} = (\boc^\alpha,\bol \boc^\beta) J_\beta.
\ees
}
\end{theorem}

\begin{proof}

Consider eq.(\ref{eq:gensol}), with $\lambda_\alpha = J_\alpha$ and $\lambda^\mu_\ast = J^\mu_\ast$, and replace in the bilinear form $\sigma = (\boj, \bol \boj)$:
\bea
\sigma = L^{\alpha\beta}J_{\alpha} J_\beta + M_{\mu\nu} J^\mu_\ast J^\nu_\ast + 2 H^\alpha_\mu J_\alpha J^\mu_\ast \label{eq:linregcom}
\eea
where we defined
\bes
M^{\mu\nu} & = & (\boe^\mu_\ast,\bol \boe^\nu_\ast)   \\
 H_\mu^\alpha & = & (\boc^\alpha,\bol\boe_\mu^\ast)  
\ees
It's simple to derive $A^\alpha = L^{\alpha\beta} J_\beta + H^\alpha_\mu J^\mu_\ast$. Completing the square: 
\be
\sigma = L^{-1}_{\alpha\beta} A^\alpha A^\beta + (M_{\mu\nu} - H_\mu^\alpha L_{\alpha\beta}^{-1} H^\beta_\nu)  J^\mu_\ast J^\nu_\ast .
\ee
Since $A^\alpha$ and $J^\mu_\ast$ are independent, setting  all affinities to zero yields the entropy production for detailed balanced systems, which after the previous theorem and Eq.\,(\ref{eq:detballin}) is easily seen to be $\sigma = (\mathbf{L}_\ast^{-1})_{\mu\nu} J^\mu_\ast J^\nu_\ast$. Since the latter is a nondegenerate bilinear form, we can identify the matrix between parenthesis with $\mathbf{L}_\ast^{-1}$.\end{proof}

This expression for the entropy production is simple and sutble; it further supports the point of view that the external currents and forces are good macroscopic nonequilibrium quantities which the observer controls. 

One physically-motivated application of Schnakenberg's macroscopic observables in the linear regime was proposed by the author \cite{polettini}, who proved that if affinities are held fixed through Lagrange multipliers while minimizing  the entropy production, the steady state is attained. Hence affinities are the correct macroscopic constraints for the minimum entropy production principle, which in one particularly suitable wording \cite{klein} asserts that
\begin{quote}
\textquotedblleft the steady state is that state in which the rate of entropy production has the minimum value consistent with the external constraints which prevent the system from reaching equilibrium\textquotedblright .
\end{quote}
Formula (\ref{eq:complete}) allows a straightforward derivation. Variation of $\sigma$ at fixed $A^\alpha, \forall \alpha$ yields
\bea 
 \frac{\delta \sigma}{\delta J_\ast^\mu} =  2 (\mathbf{L}^{-1}_\ast)_{\mu\nu} J_\ast^\nu = 0 \label{eq:minep} .
\eea
Hence we obtain $J^\nu_\ast = 0$, which characterizes the steady state. The dual proposition, which we discuss in the conclusions, follows in the same manner from  (\ref{eq:minep}), with the external currents replaced by the external affinities.

\section{\label{em}Electromagnetism on a network}

An important notion of duality against which to compare ours is the electromagnetic (EM) duality. We refer here to C. Timm's work on master equations \cite{timm}.

Let's think of $\rho$ as a charge density.
In order to make the overall network neutral we introduce a supplementary vertex ``$\infty$'', charged $\rho_\infty = - \sum_{v} \rho_v$. All graph-theoretical notation will refer to this extended graph, which can be further made into a two-dimensional cell complex by introducing a collection $P \supseteq C$ of plaquettes \cite{ddf}. Choose a conventional clockwise/counterclockwise orientation for each plaquette $p$ and define the boundary (curl) operator
\bea
(\partial \times)_{e}^p = \left\{
	\begin{array}{ll}  +1, & \mathrm{if} ~ e  \da \rcirclearrowdown p,  ~ e  \ua \lcirclearrowdown p 
	\\
	-1 , &  \mathrm{if} ~  e  \ua \rcirclearrowdown p,  ~ e  \da 	\lcirclearrowdown p
	\\
	0 , & \mathrm{elsewhere} 
	\end{array} \right.  .   \eea
Boundaries of plaquettes (columns of $\partial \times$) are cycles, hence $\partial (\partial \times) =0$, which translates into the well-known fact that the divergence of the curl vanishes.

Introduce an electric field $E_e$ over edges and a magnetic field $B_p$ over plaquettes. The electric field is required to satisfy Gauss's law $\partial \boE= \rho$. Taking the time derivative, we have $\partial (\dot{\boE} + \boj)=0$, from which it follows that $\dot{\boE}+\boj$ is a linear combination of cycles,
\bea
\boj = - \dot{\boE} + \mathscr{B}_\alpha \boc^\alpha =  -  \dot{\boE} + \partial \times \boB, 
\ee
where in the r.h.s. we imposed the Amp\`ere-Maxwell Law. Since $\boc^\alpha$ is a complete set of cycles, there exists an $|P|\times |C|$ matrix $\boeta$ such that $(\partial \times)^p  = \eta^p_{\alpha} c^\alpha$, so that $\mathscr{B}_\alpha  = \eta^p_\alpha B_p$. Further impose Faraday's Law $(\partial \times)^T \boE + \dot{\boB} =0$, and apply $\boeta$:
\bea
(\boc^\alpha,\boE) = - \dot{\mathscr{B}}_\alpha.  
\ee
It follows that any two combinations of plaquettes which share the boundary enclose a volume across whose boundary the magnetic flux is zero (Gauss's Law). Hence only $|C|$ out of $|P|$ magnetic field values are independent.

As entropy production it is reasonable to elect the total energy flux
\bea
\sigma 
= (\boE,\dot{\boE}) + \sum_p B_p \dot{B}_p = -(\boj,\boE)  = (\boE,\dot{\boE}) + \mathscr{B}_\alpha \dot{\mathscr{B}}^\alpha   
\ee
where we applied Faraday's Law, transposed the curl operator, and used Amp\`ere's Law to get the 
second identity (Integrated Poynting's Theorem). The third displays a simple dependence on the boundary values of the magnetic field. Our theorem can now be applied, yielding
\bea
\sigma = J_\alpha \dot{\mathscr{B}}^\alpha   - J_\ast^\mu \mathscr{E}^\ast_\mu  
\ee
where $\mathscr{E}^\ast_\mu$ is the electric field along cochord $\boe_\mu^\ast$.  By  (\ref{eq:sourceset}), $J_\ast^\mu$ is (minus) the time-derivative of the charge in $\texttt{s}(\boc_\ast^\mu)$. Hence under graph duality and $\boj \leftrightarrow \boE$ one obtains
\bea
\mathscr{E}_\mu^\ast \leftrightarrow J_\alpha, \quad \mathscr{B}^\alpha \leftrightarrow \rho^\mu + const. 
\eea
The electric field is mapped to the source of the magnetic field and vice versa. Thus the example further supports the interpretation of duality as reversing the role of system and environment. Although, notice that the dynamical evolution is not respected: only Kirchhoff's and Faraday's ``structure'' equations are dual to each other. The Lagrangian (see  \cite{timm}) turns out not to be self-dual. This is an important difference between sys./env. and EM duality, which is dynamical. Moreover, the former is 2-dimensional, while the latter, restricted to sourceless cases or requiring magnetic charges, involves the Hodge machinery in 3 dimensions. Contrary to standard EM duality, in ours divergencelessness of the magnetic field is an essential feature rather than an obstruction to duality.

\section{\label{concl}Discussion and conclusions}

Duality comes in many flavors in physics. Among the first that one encounters:  the duality between vectors ---velocities--- and linear forms ---momenta; the Legendre transform which maps the Lagrangian into the Hamiltonian, pivoting on the bilinear form $\sum_i \dot{q}_i p_i$; the electromagnetic duality, which is the archetypical physical counterpart of Hodge's geometrical theory of differential forms; the electro-technical duality between resistances and condensators, parallel and series reduction, voltage and current laws \cite{iyer}. The one that we put forward descends from the latter, that we generalized to nonlinear regimes, where Ohm's law does not necessarily hold; but it also resonates with each other of the above. While the reference physical situation is that of a thermodynamic system in the framework of the nonequilibrium statistical mechanics of master equation systems \cite{zia,seifert}, we cast our propositions in a very general form. In fact, they can be applied to any lattice theory which has a couple of conjugate variables.

Duality can only be realized on planar graphs. Although, nonequilibrium observables behave ``as if'' there always existed some dual graph. In a fascinating work \cite{mckee}, McKee attempts a generalization, finding a correspondence of graph duality with logical duality between the universal and existential quantifiers ($\forall$ and $\exists$) under the involutive action of negation ($\neg$). In the prologue he comments that ``some optimists see them [dualities] as mechanically doubling the number of results of a theory''. We join the optimists, claiming that for every proposition that is true of steady states, there exists a dual proposition regarding detailed-balanced systems, regardless of the technical possibility to draw a dual graph. One explicit example is the following minimum entropy production principle:
\begin{quote}
\textquotedblleft detailed balanced systems are those systems for which the rate of entropy production has the minimum value that is consistent with the fixed inflowing currents which prevent them from reaching a stationary state\textquotedblright ,
\end{quote}
which is the dual proposition to the one exposed in Sec.\,\ref{linear}.

The application to network electromagnetism highlights that duality only works for kinematical states,  \emph{viz.} instantaneus snapshots of the system.  So, for example, by ``steady'' we mean that Kirchhoff's law is satisfied, not persistence in time. This is one important limitation that one will have to take care of when considering, for example, markovian evolution: by no means do we claim that duality maps  master equations into dual master equations. As to the other important limitation, namely planarity, we already discussed how it can be formally overcome with matroids and conceptually regarded as accidental. However, from a mathematical point of view, there is another way out, based on the possibility to embed any graph on an orientable closed surface of high-enough genus. Two-dimensional dualization can then be performed on such a surface in exactly the same way. We do not discuss this possibility here, but let us just hint at some of its features. While the number of cycles and cocycles is not affected, the number of faces, hence of dual cocycles and dual cycles, will change according to Euler's formula. It is simple to foresee that Theorem \ref{theorem1} will hold unchanged, but its interpretation will have to be accordingly modified, accounting for a number of global currents and of topological phases, such as those which were taken in consideration by Jiang and the Qians \cite{qiansbook} in their geometrical characterization of circulation on manifolds. In that cases, duality will only hold locally. This approach, whilst much more concrete than matroids, is doomed to become impracticable when one deals with lattices of more than two dimensions, in a thermodynamic limit. Beyond two dimensions, there is a gap between the mathematical realization of duality, which suffers from great limitations (abstractness, in the case of matroids, and excessive complication in the case of surfaces) and the propositional reach of the theory, which seems to be completely independent of the possibility to visualize duals.

\begin{figure}
\hspace{3cm} \includegraphics[scale=0.5]{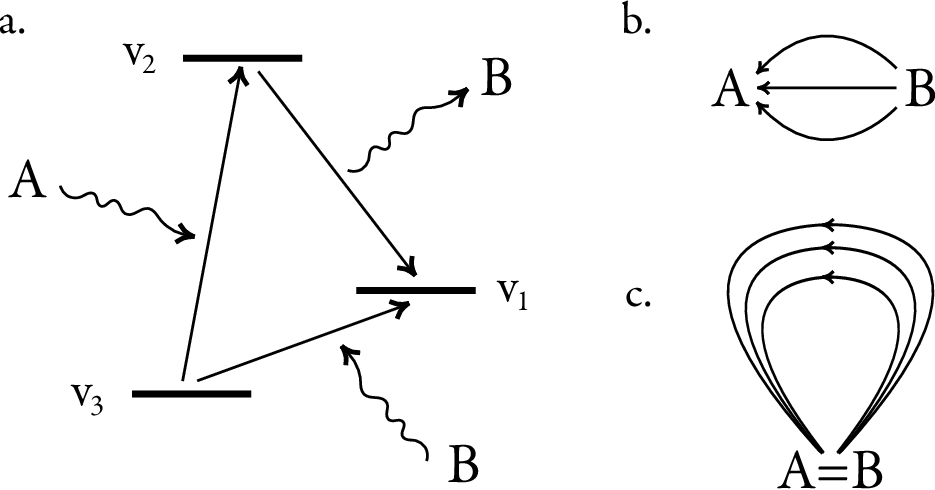}
\caption{\label{fig2}(a.) Transitions between states  due to absorption and emission from two reservoirs. (b.) Steady state heat flux between reservoirs. (c.) One reservoir with internal fluxes stimulated by the system's nonsteady configuration.}
\end{figure}

To conclude, let us linger on the 3-state example, in the attempt to provide a more intuitive grasp on the physics of duality. Suppose the labels $v_i$ of the example graph are energy levels of an open system, which can emit and absorb energy from the environment Fig.\,\ref{fig2}). The onset of a NESS might be due to the interaction with two thermal baths \cite{esposito}, whose inverse temperatures $\beta_A$ and $\beta_B$ label the states of the dual system, with $\beta_A > \beta_B$. Suppose that transitions 2 and 3 are exclusively due to the interaction with B, while transition 1 is exclusively due to the interaction with A. The ratio of emission and absorption  rates is given by $w_{e_1}/w_{-e_1} = \exp \beta_A (v_2-v_3)$, and similarly for the others, yielding as macroscopic affinity
$A_1 = (\beta_A - \beta_B) (v_2-v_3)$. In a nonequilibrium steady state, with current $j_1=j_2=j_3 = J_1$, one transition yielding an amount of energy $v_2-v_3$ happens on average every $|J_1|^{-1}$ seconds, while in the same time two transitions, which absorb respectively amounts of energy $v_2-v_1$ and $v_1-v_3$, are stimulated by the interaction with reservoir B.  It takes shape a picture where to a steady state there corresponds a nonsteady flow of energy from the hotter to the colder bath:
\begin{quote}
  non  steady sys. $\rightarrow$ nonsteady env.
  \end{quote}
 Whilst purely speculative, this interpretation is consistent with the physical intuition that NESSs are determined by a transient environmental behavior \cite{janotta}. 
Vice versa, a detailed-balanced flow arises when there is no temperature gradient, $\beta_A = \beta_B$, in which case we only resolve one reservoir. At equilibrium, because of steadiness and detailed balancing, as many emitting and absorbing transitions occur. However, fluxes within the system determine  a non-null flow of currents in the bath. The latter, being a 1-state system, is necessarily in a steady state. Hence the system's state plays the role of external force which causes internal fluxes to flow within the environment:
\begin{quote}
steady env. $\rightarrow$ det. bal. nonsteady sys.
\end{quote} 
This is nothing but the logical negation of the above proposition, hence its dual under transposition of the material implication symbol ($\rightarrow$), in the spirit of McKee's logical interpretation of duality.

Despite of its simplicity, the example is rather clumsy and only vaguely illustrative: system and environment do not play mirror roles, for which reason we were not able to draw the inverse implications. However, the qualitative principle seems to be robust.  It is quite remarkable  that graph duality finds a similar interpretation also in mechanical engineering \cite{shaiA,shaiB}, where the statics of  structures and machines  and  their first order kinematics are related to dual properties of their design. Thus there seems to be a vast variety of systems to which duality might apply: it is the author's opinion that the development of a complete statistical model which displays duality between enviromental and internal degrees of freedom would be a major advance.

\section*{Acknowledgments}

The author warmly thanks A. Maritan and D. Andrieux for discussion,  M. Esposito and M. Dalmonte for helping out with the first drafts. The research was partly supported by the National Research Fund Luxembourg in the frame of the AFR Postdoc Grant 5856127.

\end{document}